\documentclass[twocolumn,prl,aps,showpacs]{revtex4}

\usepackage{graphicx}%

\usepackage{dcolumn}
\usepackage{amsmath}

\usepackage[caption=false]{subfig}

\newcommand{\ed}{\end{document}}

\newcommand{\ice}[1]{\relax}

\newcommand{\as}{a_s}

\newcommand{\als}{\alpha_s}
\newcommand{\beq}{\begin{equation}}

\newcommand{\ba}{\begin{array}}

\newcommand{\ea}{\end{array}}

\newcommand{\eeq}{\end{equation}}

\newcommand{\bea}{\begin{eqnarray}}

\newcommand{\eea}{\end{eqnarray}}

\newcommand{\nnb}{\nonumber}

\begin{document}

\title{
Complete QCD Corrections to   Hadronic $Z$--Decays 
in Order $\alpha_s^4$
 }     

\author{P.~A.~Baikov}
\affiliation{
Skobeltsyn Institute of Nuclear Physics, Lomonosov Moscow State University\\
1(2), Leninskie gory, Moscow 119234, Russian Federation
        }

\author{K.~G.~Chetyrkin}\thanks{{\small Permanent address:
Institute for Nuclear Research, Russian Academy of Sciences,
 Moscow 117312, Russia.}}
\author{J.~H.~K\"uhn}
\affiliation{Institut f\"ur Theoretische Teilchenphysik,
  Universit\"at Karlsruhe, D-76128 Karlsruhe, Germany}

\author{J.Rittinger}
\affiliation{Institut f\"ur Theoretische Teilchenphysik,
  Universit\"at Karlsruhe, D-76128 Karlsruhe, Germany}

\begin{abstract}
\noindent

Corrections of order $\alpha_s^4$ for
the axial singlet contributions for the decay rate of the $Z$-boson
into hadrons are evaluated in  the limit of the heavy  top quark  mass.
Combined with recently finished ${\cal O}(\alpha_s^4)$ calculations
of the non-singlet corrections, the new results directly  lead us
to  the first {\em complete}  ${\cal O}(\alpha_s^4)$ prediction for
the total hadronic decay rate of the Z-boson. 

The  new ${\cal O}(\alpha_s^4)$  term in Z-decay rate   lead to a
significant stabilization of the perturbative series, to a reduction
of the theory uncertainty in the strong coupling constant $\alpha_s$,
as extracted from these measurements, and to a small shift of the
central value.

\end{abstract}


\maketitle



The precise determination of the $Z$-boson decay rate into hadrons at
LEP \cite{Alcaraz:2007ri} has led to one of the most precise determinations of
the strong coupling constant $\alpha_s$. From the experimental side, in
view of the fully inclusive nature of this measurement, the result is
fairly robust, in particular since it is insensitive to simulations of
the hadronic final state. Hence the error is essentially dominated by
the statistical uncertainty. From the theory side the advantage of the
measurement is its high energy, and as a result, the irrelevance of
nonperturbative and power-law suppressed terms. The smallness of
$\alpha_s$ at high energies then leads to a rapid decrease of higher
order corrections in the perturbative series and, correspondingly, to
a significant reduction of the theory error. 

A variety of methods has been suggested to estimate the remaining
uncertainty in the theory prediction. Using the last calculated term
is probably the most conservative approach, varying the
renormalization scale $\mu$ within an energy range characteristic for
the problem (e.g. $M_Z/3 < \mu < 3\,M_Z$) is frequently used, albeit
with considerable ambiguity in the actual choice of the region of the
$\mu$-variation. In order to reduce the theoretical uncertainty in the
extraction of $\alpha_s$ to a level significantly smaller than the
experimental one (which amounts to $\pm 0.0026$ at present
\cite{Alcaraz:2007ri}), the knowledge of the corrections of ${\cal
O}(\alpha_s^4)$ is necessary. At the same time this calculation opens
the window for a considerable improvement in the
$\alpha_s$-determination at $Giga\,Z$, the project of a
high-luminosity linear collider operating at the $Z$-resonance (see
e.g. \cite{Winter:2001av}, where a precision of 0.0005 to 0.0007 has
been advertised).  The dominant part of the $\alpha_s^4$-corrections,
the ''non-singlet''-piece, has been evaluated in
\cite{Baikov:2008jh}. This has lead to a slight shift of the central
value of $\alpha_s$ upwards from $0.1185\pm 0.0026$ to $0.1190\pm 0.0026$
\cite{Baikov:2008jh} and a reduction of the theory error far below the
error of 0.0026 from experiment. However, as noted already in
\cite{Baikov:2008jh}, for a complete evaluation of the decay rate in
${\cal O}(\alpha_s^4)$ an additional set of corrections, namely those
for the ``singlet'' contributions, is required. For the axial current
correlator these start at ${\cal O}(\alpha_s^2)$
\cite{Kniehl:1989qu,Kniehl:1989bb}, for the vector correlator at
${\cal O}(\alpha_s^3)$. Both of them are presently known to third
order in $\alpha_s$ only
\cite{Gorishnii:1990vf,Surguladze:1990tg,Chetyrkin:1993jm,Chetyrkin:1993ug,Larin:1993ju}. 
Hence, for a completely consistent ${\cal O}(\alpha_s^4)$
extraction of the strong coupling the extension of these results by one
order in $\alpha_s$ is required.

Before describing this calculation in detail, let us briefly recall 
the basic structure of QCD corrections to the correlator of the
electromagnetic and the neutral current, respectively, their similarities 
and their main differences. After splitting off inessential kinematic
factors, the absorptive part of the current-current correlator of the
electromagnetic current 
is expressed by the familiar $R$-ratio 
\begin{equation}
R^{\rm  em}=3\Big[\sum_f q_f^2 r_{\rm NS}^V + (\sum_f q_f)^2 r_{\rm S}^V\Big]~, \label {Rem}
\end{equation}
where $r_{\rm NS}^V$ and $r_{\rm S}^V$ stand for the (numerically dominant) non-singlet
and the singlet part respectively. The corresponding decomposition
for the correlator of the neutral current 
involves the following four terms
\begin{equation}
R^{\rm  nc}=3\Big[\sum_f\!v_f^2 r_{\rm NS}^V + \big(\sum_f\!v_f\big)^2 r_{\rm S}^V
+\sum_f\!a_f^2 r_{\rm NS}^A + r_{\rm S;t,b}^A\Big]~,
\end{equation}
with $v_f\equiv 2I_f-4q_fs_W^2 $, $a_f \equiv 2I_f$ and $s_W$ defined as effective weak mixing angle. 
Here all but the top quark are assumed to be massless.  

(Mass corrections to both vector and axial vector correlator due to
other massive quarks are dominated by the bottom quark and can be
classified by orders in $m_b^2/M_Z^2$ and $\alpha_s$.
Up to  ${\cal O}(\als^2 m_b^2/M_Z^2)$ and
${\cal O}(\als^2 m_b^4/M_Z^4)$ they can be found in 
\cite{Chetyrkin:1996ia}, as well terms of  order 
$\als^2 m_b^2/M_Z^2$ (const + $\log\   m_b^2/M_Z^2$)
and  $\als^2 m_b^2/M_t^2$ (const + $\log\  m_b^2/M_Z^2$)
that arise  from  axial vector singlet contributions. Terms  of order 
$\als^3 m_b^4/M_Z^4$ and $\als^4 m_b^2/M_Z^2$ can be found in  
\cite{Chetyrkin:2000zk} and
\cite{Baikov:2004ku} respectively.
Corrections of order $\als^2 m_Z^2/m_t^2$ and  $\als^3 m_Z^2/m_t^2$ from 
singlet and non-singlet terms are known from \cite{Kniehl:1989qu,Kniehl:1989bb,Chetyrkin:1993tt} and \cite{Larin:1994va} respectively. 
These are important for the
actual $\alpha_s$-determination, but will not be discussed further in
the present paper.)

From the prefactors of the
non-singlet contributions in electromagnetic, vector and axial
correlator it is evident that different quark flavours contribute
incoherently, hence additive to the rate. Thus their contribution is significantly
enhanced in comparison with the singlet terms where amplitudes from
different flavours interfere destructively, with prefactors $(\sum_f
q_f)^2$ and $(\sum_f v_f)^2$ for the electromagnetic and neutral
current respectively.

\begin{figure}[b!]
\subfloat[]{
\includegraphics[width=.25\linewidth]{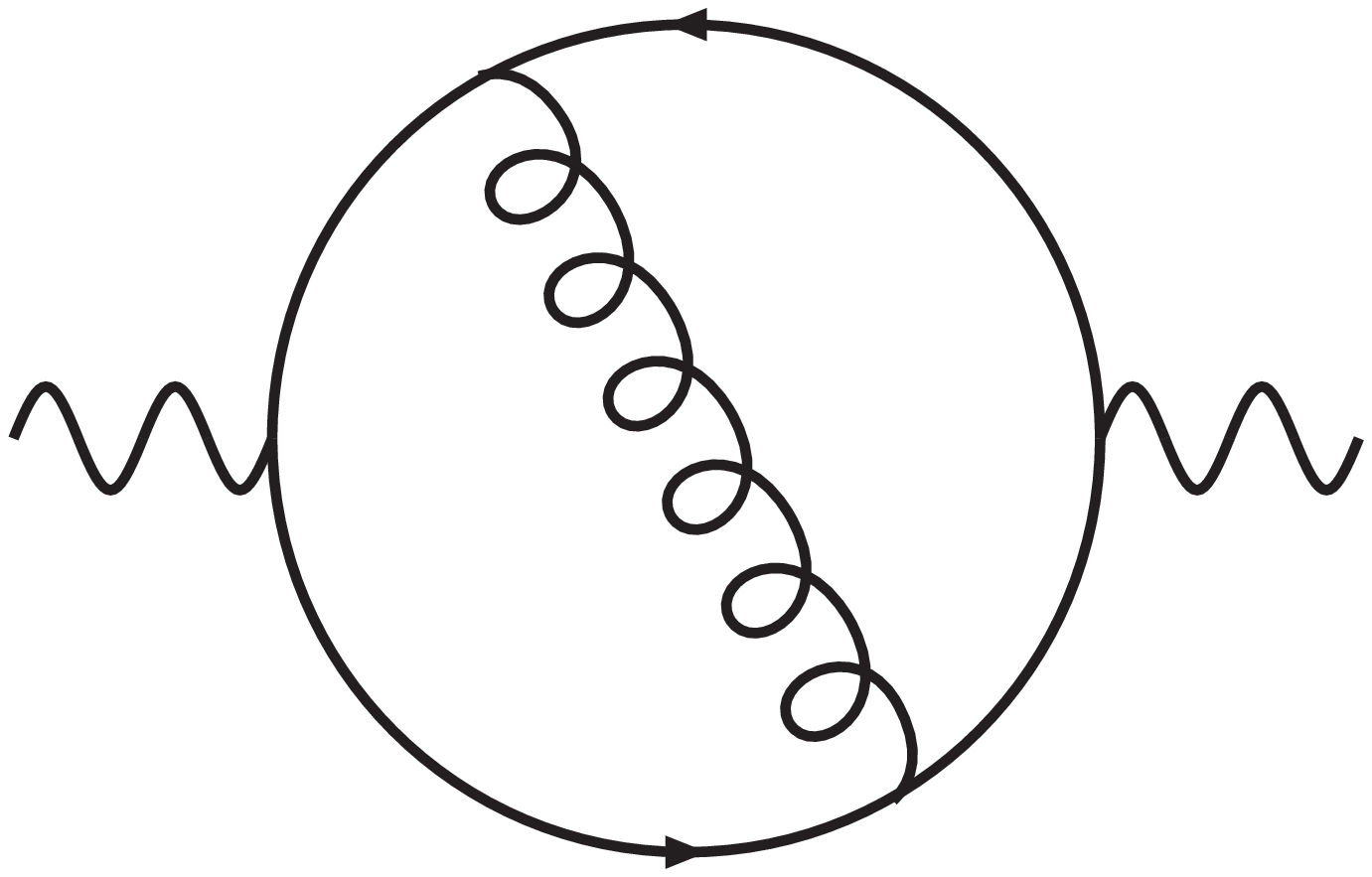}
\label{PropNS}
}
\subfloat[]{
 \includegraphics[width=.3\linewidth]{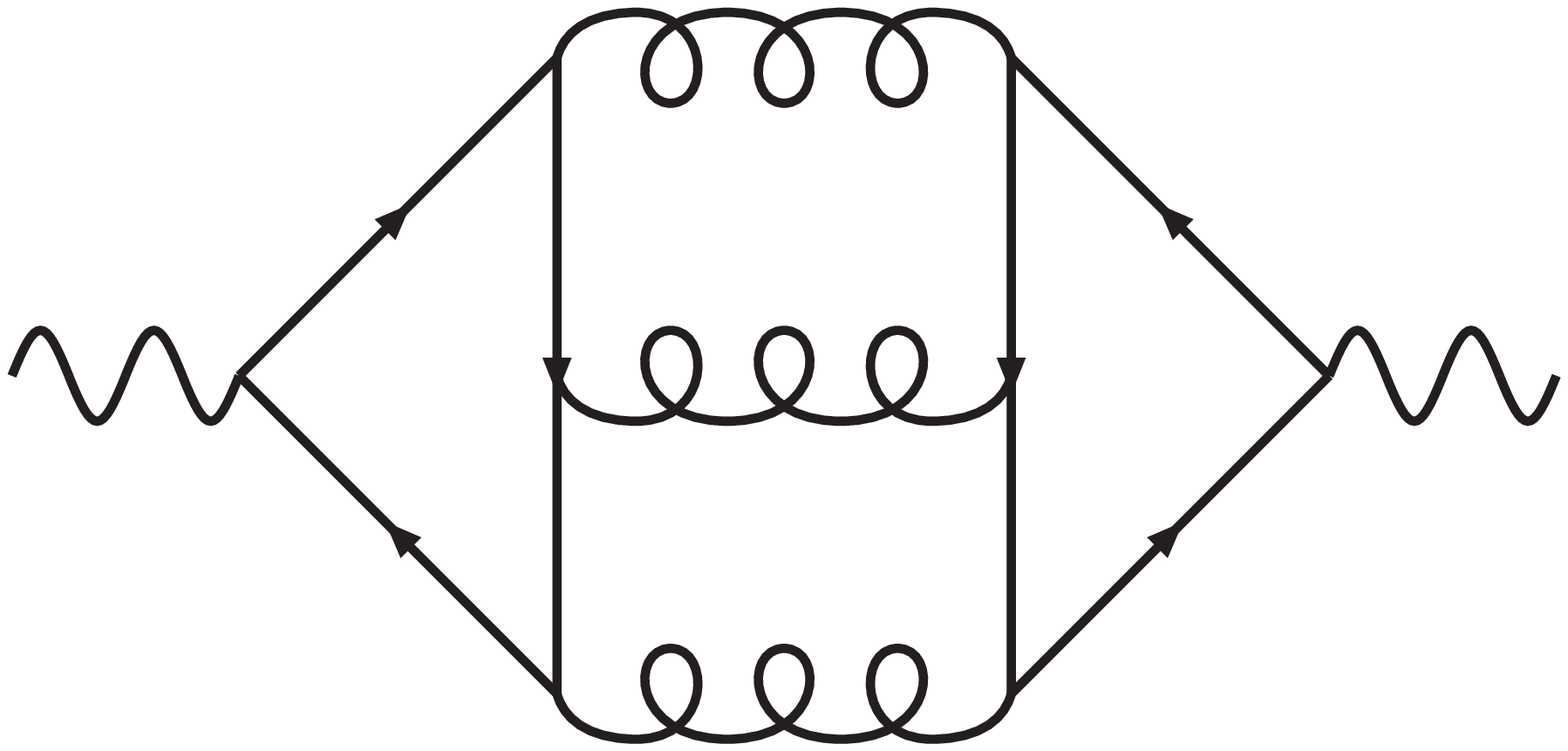}
\label{PropSV}
}
\subfloat[]{
 \includegraphics[width=.3\linewidth]{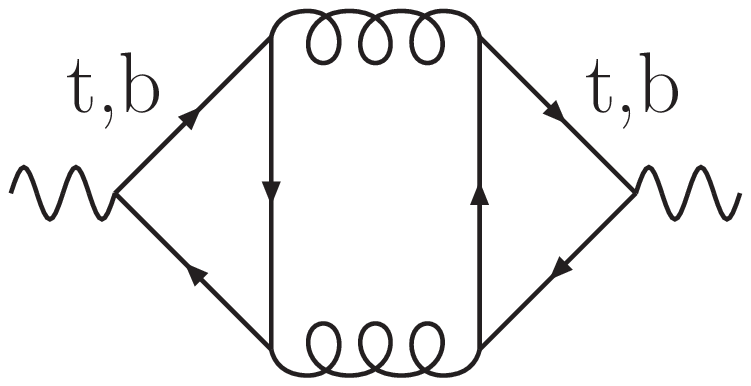}
\label{PropSA}
}
\caption{Different contributions to $r$-ratios: (a) non-singlet, (b) vector singlet and (c) axial vector singlet.}
\end{figure}

Non-singlet contributions are present at the parton level
and the QCD corrections are known in second \cite{Chetyrkin:1979bj}, third
\cite{Gorishnii:1990vf,Surguladze:1990tg} and fourth \cite{Baikov:2008jh} order in $\alpha_s$. In terms of
Feynman diagrams, non-singlet contributions are characterized by the
fact that one quark loop connects the two external currents (Fig. \ref{PropNS}).
In the absorptive part of this fermion loop no top quark is present due to kinematic reasons,
whence the non-singlet functions are identical 
$r_{\rm NS}^V=r_{\rm NS}^A\equiv r_{\rm NS}$. 

In the case of singlet contributions of the vector current the two currents couple to two
different quark loops (Fig. \ref{PropSV}) requiring a three-gluon intermediate
state. Correspondingly the leading term is of ${\cal O}(\alpha_s^3)$
and has been obtained already long time ago \cite{Gorishnii:1990vf,Surguladze:1990tg}. The NLO
corrections to this   result are of ${\cal O}(\alpha_s^4)$. They serve
to soften the strong scale dependence of the ${\cal O}(\alpha_s^3)$  result, stabilize the theory prediction and will be the
subject of this paper.

The situation is different in the case of the singlet axial vector 
current correlator. The axial couplings of the two members of an isospin 
doublet are opposite equal. Hence their singlet contributions vanishes, if the
corresponding quark masses are equal. This approximation is valid for
the two lightest quark doublets. The only remaining contribution
originates from the combination of bottom and top quarks with their 
specific mass hierarchy  $m_b^2<<M_Z^2<<m_t^2$ (Fig. \ref{PropSA}). In this case
the contribution starts at ${\cal O}(\alpha_s^2)$ and is further
enhanced by the ''large'' logarithm $\log (m_t^2/M_Z^2)$ 
\cite{Kniehl:1989qu,Kniehl:1989bb}. Corrections of ${\cal O}(\alpha_s^3)$
have been calculated in \cite{Chetyrkin:1993jm,Chetyrkin:1993ug,Larin:1993ju}, those of  ${\cal O}(\alpha_s^4)$
will be the subject of this paper.

The evaluation of the NLO terms of $r_V^{\rm S}$ requires the
calculation of the absorptive parts of five-loop diagrams with
massless propagators which, with the help of some complicated
combinatorics based on the $R^*$-operation \cite{Chetyrkin:1984xa}, can be boiled down
to the calculation of  four-loop propagator
diagrams. The  latter have been computed  
via reduction to 28  master integrals, based on evaluating  
sufficiently many terms of the $1/D$ expansion \cite{Baikov:2005nv} of
the corresponding coefficient functions \cite{Baikov:1996rk}. 
This direct procedure required huge computing resources and
was performed using a parallel version \cite{Tentyukov:2004hz} of FORM
\cite{Vermaseren:2000nd}. The master integrals are reliably  known from
\cite{Baikov:2010hf,Smirnov:2010hd,Lee:2011jt}. 
The details of the calculation, the results in analytic form and their 
relation to the Gross-Llewellyn Smith sum rule will be given in  \cite{Baikov:2012}.

The evaluation of the NNLO terms of $R_{\rm S;t,b}^A$ involves again
absorptive parts of five-loop diagrams with massless propagators,
however, in addition also absorptive parts of four-loop diagrams
combined with one-loop massive tadpoles, etc. down to one-loop
massless diagrams together with four-loop massive tadpoles. The latter
have been computed with the help  of the  Laporta algorithm
\cite{Laporta:2001dd} implemented in Crusher \cite{crusher}. 
The methods employed in our calculations, together with the results
 will be described in more detail in
\cite{Baikov:2012}.
\ice{, together with the results expressed in terms of
group theoretical coefficients and transcendental numbers.
}

The result is valid
in the limit $M_Z^2 \ll 4\,M_t^2$, an excellent approximation as evident
from the lower orders.
The relative importance of the various terms
is best seen from the results for the various $r$-ratios introduced
above, expressed in numerical form
\begin{align}
 r_{\rm NS}=&1 + \as + 1.4092\,\as^2 - 12.7671\,\as^3 - 79.9806\,\as^4~, \nnb \\
 r_{\rm S}^V=&-0.4132\,\as^3 - 4.9841\,\as^4~, \nnb \\
 r_{\rm S:t,b}^A =\, &  (-3.0833+l_t)\,\as^2  \\ &+ (-15.9877+3.7222\,l_t+1.9167\,l_t^2)\,\as^3 \nnb \\
		 &~\hspace{-1cm} +( 49.0309-17.6637\,l_t+14.6597\,l_t^2 + 3.6736\,l_t^3)\,\as^4  ~, \nnb
\end{align}
with $\as=\alpha_s(M_Z)/\pi$ and $l_t=\ln(M_Z^2/M_t^2)$. Since all
three $r$-ratios are separately scale invariant, the corresponding results
for a generic value of $\alpha_s(\mu)$ can easily be reconstructed. 
Using for the pole mass $M_t$ the value 172 GeV the axial singlet contribution is given
in numerical form by
\begin{align}
 r_{\rm S;t,b}^A = -4.3524\,\as^2 - 17.6245\,\as^3 + 87.5520\,\as^4~.
\end{align}
Collecting now all QCD terms, the decay rate of the $Z$-boson into hadrons 
can be cast into the following form
\begin{align}
 \Gamma_Z=\Gamma_0\,R^{\rm nc}=\tfrac{G_F\,M_Z^3}{24\pi\sqrt{2}}\,R^{\rm nc}~.
\end{align}
Here all electroweak corrections are assumed to be collected in the
prefactor $\Gamma_0$, and the forementioned mass corrections are
ignored as well as electroweak and mixed QCD-electroweak corrections 
\cite{Czarnecki:1996ei,Harlander:1997zb,Czarnecki:1996ac}. Thus the $R$-ratio is now known up to $\mathcal{O}(\as^4)$
\begin{align}
 R^{\rm nc}=&20.1945 + 20.1945\,\as \nnb \\ &+ (28.4587-13.0575+0)\,\as^2 \nnb \\ &+(-257.825-52.8736-2.12068)\,\as^3 \nnb \\ &+(-1615.17+262.656-25.5814)\,\as^4~,
\end{align}
with $s_W^2=0.231$. The three terms in the brackets display separately non-singlet, axial singlet and vector singlet contributions.

Let us now evaluate the impact of the newly calculated terms on the
$\alpha_s$-determination from $Z$-decays. Following our approach for
the non-singlet terms (where a shift $\delta\alpha_s=0.0005$ had been
obtained \cite{Baikov:2008jh}, consistent with an analysis \cite{Quast}  based on
results of the electroweak working group \cite{Alcaraz:2007ri}
and a modified interface to  \mbox{ZFITTER v. 6.42} \cite{Bardin:1999yd,Arbuzov:2005ma}
and confirmed by the G-fitter collaboration 
 [32,30,31]),
we consider the
quantity $R^{\rm nc}$ as ``pseudo-observable''. With a starting value
$R^{\rm nc}=20.9612$, if evaluated for $\alpha_s=0.1190$ and without the
$\alpha_s^4$ singlet terms, a shift $\delta\alpha_s=-0.00008$
is obtained after including the newly calculated
contributions.

\begin{figure}[b!]
\centering
\subfloat[]{
\hspace{.45cm}
\includegraphics[width=.95\linewidth,height=5.4cm]{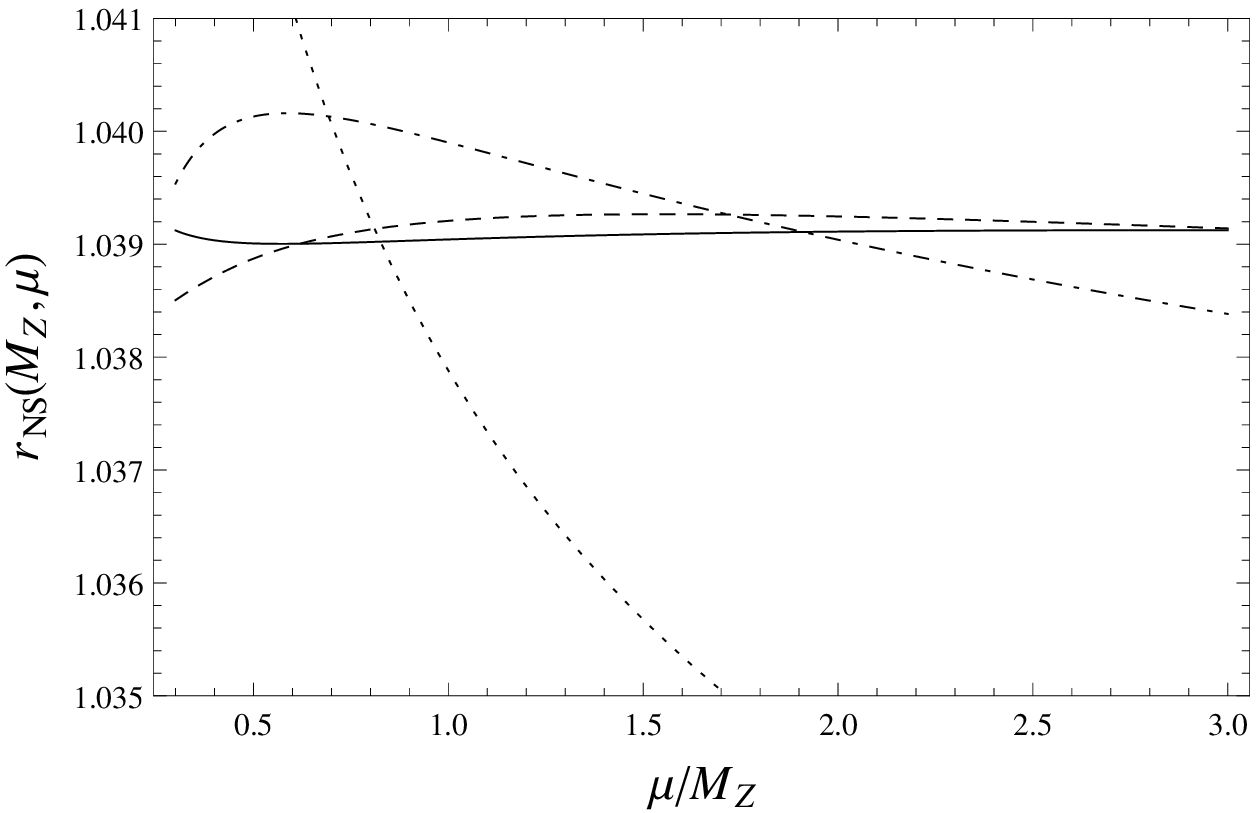}
\label{rns}
} \\
\subfloat[]{
\includegraphics[width=\linewidth,height=5.4cm]{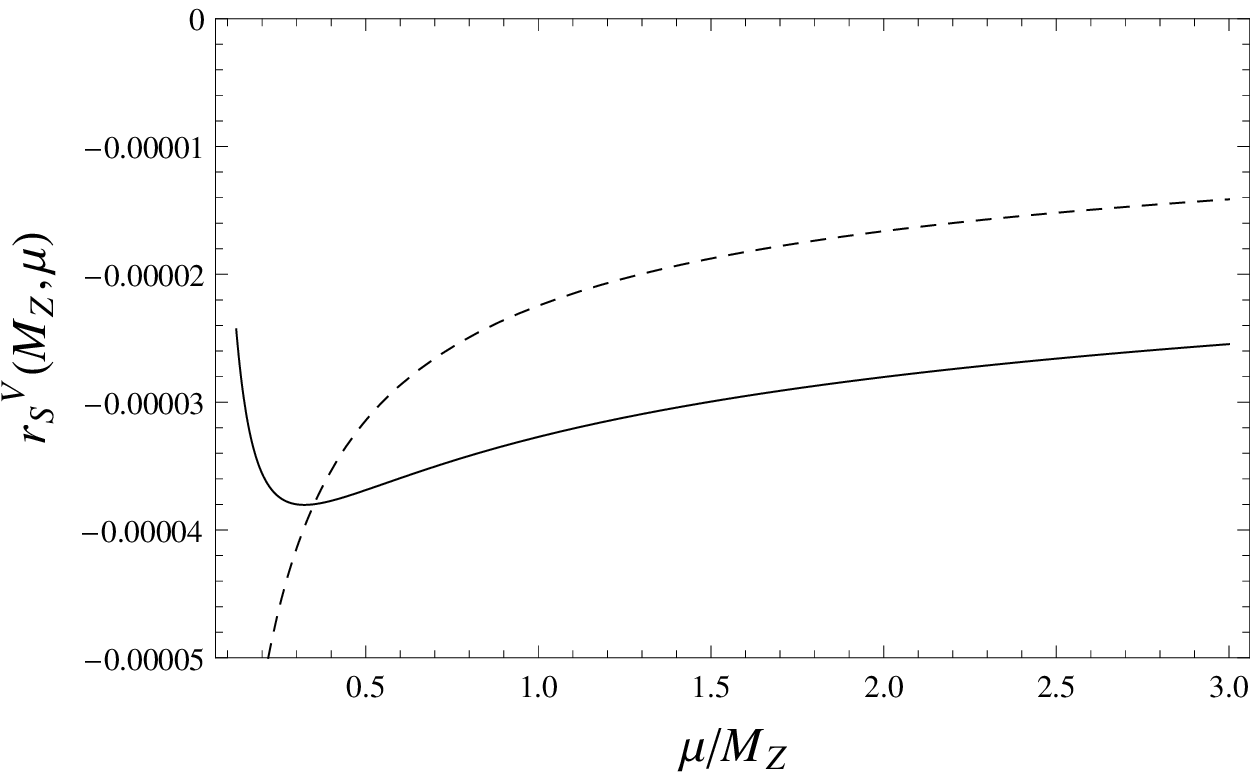}
\label{rsv}
} \\
\subfloat[]{
 \hspace{.2cm}
\includegraphics[width=.98\linewidth,height=5.4cm]{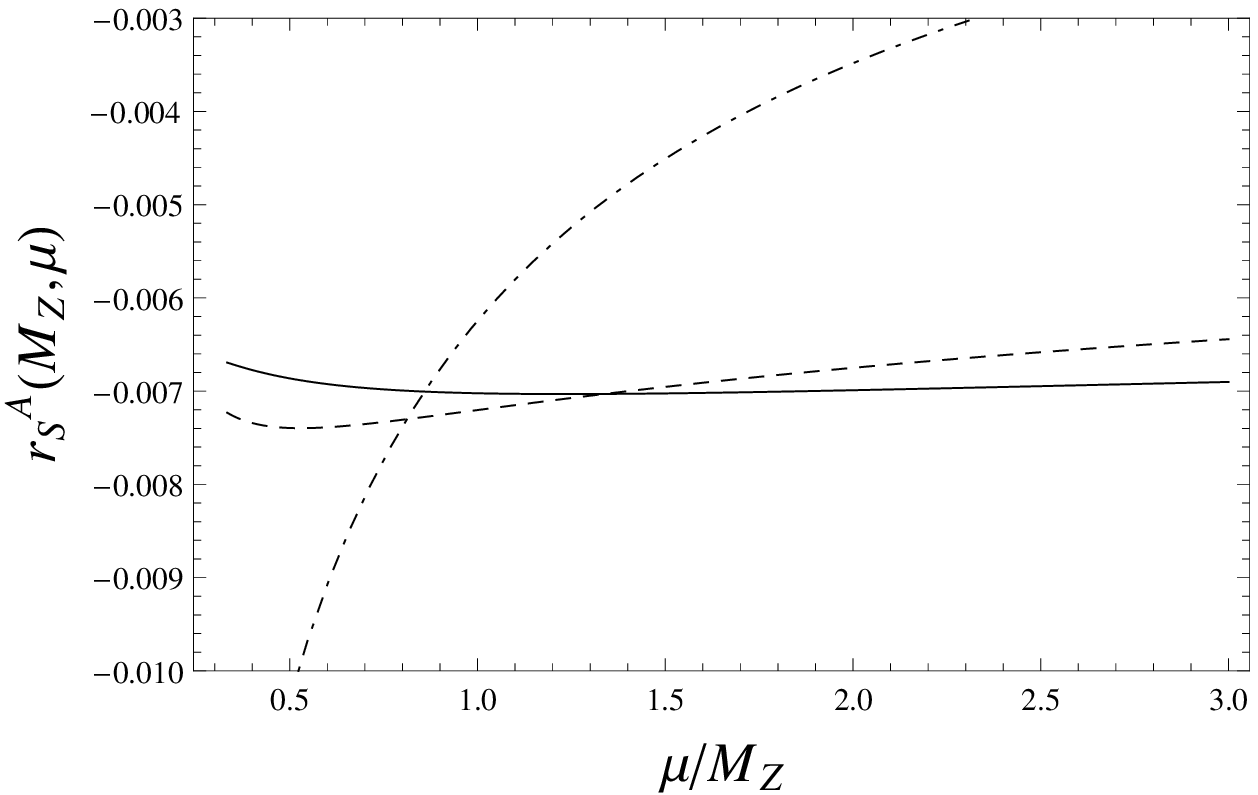}
\label{rsa}
}
\caption{Scale dependence of (a) non-singlet $r_{\rm NS}$, (b) vector singlet $r_S^V$ and (c) axial vector singlet $r^A_{\rm S;t,b}$. Dotted, dash-dotted, dashed and solid curves refer to $\mathcal{O}(\alpha_s)$ up to $\mathcal{O}(\alpha_s^4)$ predictions.
$\alpha_s (M_Z) =0.1190$ and $n_l =5 $ is adopted in all these curves.
}
\label{Plots}
\end{figure}

As discussed in \cite{Baikov:2008jh}, the non-singlet $\alpha_s^4$
term leads to a considerable stabilization of the theory prediction,
and, correspondingly, to a reduction of the theory error. A similar
statement holds true for the singlet contribution. To illustrate this
aspect, the dependence on the renormalization scale $\mu$ is shown in
Fig. \ref{Plots} for $r_{\rm NS}$, $r_{\rm S}^V$ and $r^A_{\rm S;t,b}$. The relative
variation is significantly reduced in all three cases. In particular
for the vector singlet case we observe a shift of the result by about
a factor 1.45 (for $\mu=M_Z$) and a considerable flattening of the
result. Using for example the Principle of Minimal Sensitivity (PMS)
\cite{Stevenson:1981vj} as a guidance for the proper choice of scale,
$\mu=0.3\,M_Z$ seems to be favoured, leading to an amplification of
the LO result by a factor 1.68 (if the latter is evaluated for
$\mu=M_Z$, as done traditionally). 
\ice{
On the other hand LO and NLO
predictions, if evaluated for $\mu=0.3\,M_Z$, are fairly close which
is another argument in favor of this choice. 
}

Let us assume that the remaining theory uncertainties from $r_{\rm NS}$, $r^V_{\rm S}$ and
$r^A_{\rm S;t,b}$ can be estimated by varying $\mu$
between $M_Z/3$ and $3\,M_Z$ and using the maximal variation as twice
the uncertainty $\delta r$. This leads to $\delta\Gamma_{\rm NS}=0.101$
MeV, $\delta\Gamma^V_{\rm S}=0.0027$ MeV and $\delta\Gamma^A_{\rm S}=0.042$
MeV. Even adding these terms linearly, they are far below the
experimental error of $\delta\Gamma_{exp}=2.0$ MeV \cite{Nakamura:2010zzi}.
In combination with the quadratic and quartic mass terms, which are
known to $\mathcal{O}(\alpha_s^4)$ and $\mathcal{O}(\alpha_s^3)$
respectively, this analysis completes the QCD corrections to the $Z$
decay rate.

\begin{figure}[h!]
\centering
 \includegraphics[width=\linewidth]{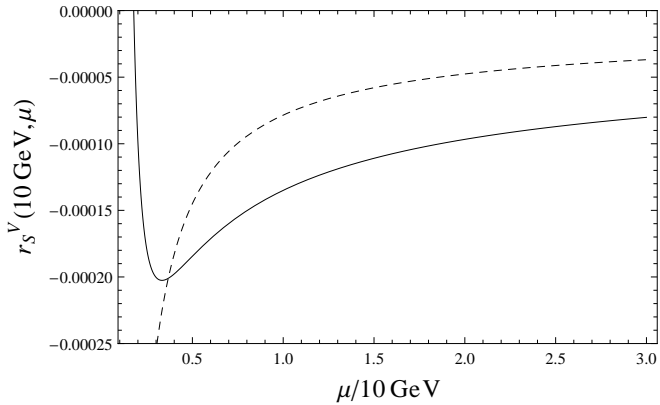}
\label{rsv10}
\vspace{-.33cm}
\caption{Scale dependence of the vector singlet $r_S^V$ around 10 GeV. Dashed and solid curves refer to $\mathcal{O}(\alpha_s^3)$ and $\mathcal{O}(\alpha_s^4)$ predictions.
$n_l =4 $ and $\alpha_s (10 \mbox{GeV}) = 0.1806$ as obtained with the use of  package RunDec \protect\cite{Chetyrkin:2000yt}
have  been assumed.
}
\label{Plot10gev}
\end{figure}

Let us also comment on the impact of the $\alpha_s^4$ singlet result
on the measurement of $R^{\rm em}$ at low energies, i.e. in the region
accessible at BESS or at  B-factories, say between 3 GeV and 10
GeV. Considering the large luminosities collected at these machines, a
precise $\alpha_s$ determination from $R^{\rm em}$ seems possible
\cite{Chetyrkin:1996tz}. In the 
low energy region only $r_{\rm S}^V$ and $r_{\rm NS}^V$ contribute. Since
$\sum_{f=u,d,s}\,q_f=0$, the singlet contribution vanishes in the
three flavour case. If we consider the region above charm and below
bottom threshold, say at 10 GeV, only $u,d,s$ and $c$ quarks
contribute, the relative weight of the $r_S^V$ in eq. (\ref{Rem}) is
given by $(\sum q_f)^2/(\sum q_f^2) = 2/5$, and thus is fairly
suppressed. At energy of 10 GeV, in the absence of open bottom quark
contribution, it seems appropriate to analyze the results in an
effective four flavour theory with 
\[
r^V_S = -0.41318 \,\as^3(\mu)  - (5.1757 + 2.5824 \ln \mu^2/s)\,\as^4(\mu)
{}.
\]
As shown in Fig. \ref{Plot10gev}, it is evident that the
scale dependence is softened in NLO. Again a scale $\mu$ around
$0.3\,\sqrt{s}$ minimizes the NLO corrections. 
\ice{
Note, that in this case LO and
NLO results differ by nearly a factor two; nevertheless the
contribution to $R^{\rm em}$ remains  small.
}

\vspace{-.23cm}
In conclusion we want to mention that all our calculations have been
performed on a SGI ALTIX 24-node IB-interconnected cluster of 8-cores
Xeon computers 
using  parallel  MPI-based \cite{Tentyukov:2004hz} as well as thread-based \cite{Tentyukov:2007mu} versions  of FORM
\cite{Vermaseren:2000nd}.  For evaluation of color factors we have used the FORM program { COLOR}
\cite{vanRitbergen:1998pn}. The diagrams have been generated with QGRAF \cite{Nogueira:1991ex}.
This work was supported by the Deutsche Forschungsgemeinschaft in the
Sonderforschungsbereich/Transregio SFB/TR-9 ``Computational Particle
Physics'', by Graduiertenkolleg 1694 
``Elementarteilchenphysik bei h\"ochster Energie und h\"ochster Pr\"azision''
 and  by RFBR grants   11-02-01196 and  10-02-00525.

We thank P. Marquard for  his friendly help with the package Crusher.

\end{document}